\documentclass[final]{svjour3}
\usepackage{graphicx}
\usepackage{rotating}
\usepackage{amssymb}
\usepackage{mathptmx}
\usepackage[numbers]{natbib}
\makeatletter
\journalname{Journal of Low Temperature Physics}
\usepackage[english]{babel}
\bibpunct{[}{]}{,}{n}{}{,}

\begin{document}

\newcommand{\hdblarrow}{H\makebox[0.9ex][l]{$\downdownarrows$}-}
\title{Development of an Organic Plastic Scintillator based Muon Veto Operating at Sub-Kelvin Temperatures for the NUCLEUS Experiment}

\author{A. Erhart$^{1,2,*}$ \and  V. Wagner$^{1}$ \and  L. Klinkenberg$^{1}$ \and \\ T. Lasserre$^{2}$ \and D. Lhuillier$^{2}$ \and C. Nones$^{2}$ \and R. Rogly$^{2}$ \and V. Savu$^{2}$ \and R. Strauss$^{1}$ \and M. Vivier$^{2}$ \\{\normalfont on behalf of the NUCLEUS collaboration}}

%\author{A. Erhart$^{1,2,*}$ \and {\normalfont on behalf of the NUCLEUS collaboration}}

\institute{$^{1}$ Physik-Department, Technische Universit\"at M\"unchen, 85748 Garching, Germany\\
$^{2}$ IRFU, CEA, Universit\'{e} Paris Saclay, 91191 Gif-sur-Yvette, France\\
$^{*}$ \email{andreas.erhart@tum.de}}

\maketitle

\begin{abstract}

The NUCLEUS experiment aims at measuring the coherent elastic scattering of nuclear reactor antineutrinos off nuclei using cryogenic calorimeters. Operating at an overburden of 3\,m.w.e., muon-induced backgrounds are expected to be dominant. It is therefore essential to develop an efficient muon veto, with a detection efficiency of more than 99\,\%. This will be realized in NUCLEUS through a compact cube assembly of plastic scintillator panels. In order to prevent a large unshielded area where the cryostat intersects the shielding arrangement without unnecessarily increasing the induced detector dead time, a novel concept has been investigated, featuring a plastic scintillator based active muon veto operating inside the NUCLEUS cryostat at sub-Kelvin temperatures. The verification of the key physical aspects of this cryogenic muon veto detector led to the first reported measurements of organic plastic scintillators at sub-Kelvin temperatures. The functionality of the principal scintillation process of organic plastic scintillators at these temperatures has been confirmed. On the basis of these findings, a disc-shape plastic scintillator equipped with wavelength shifting fibers and a silicon photomultiplier to guide and detect the scintillation light has been developed. The NUCLEUS cryogenic muon veto will be the first of its kind to be operated at sub-Kelvin temperatures.

%For this reason, an efficient muon veto with a detection efficiency of more than 99\,\% is indispensable and shall be achieved in NUCLEUS with a compact cube assembly of plastic scintillator panels.  


\end{abstract}

\section{Introduction}
% Introduction to CEvNS
The occurrence of neutral-current interactions within the Standard Model of Particle Physics implies the coupling of neutrinos to quarks through exchange of a neutral Z boson \cite{hasert1974}. In 1974, Daniel Z. Freedman concluded the existence of elastic neutrino-nucleus scattering - a process which is not limited by an energy threshold and thus predicts strong coherent effects at very low neutrino energies of $\mathcal{O}$(10\,MeV) \cite{freedmann1974}. The exploration of coherent elastic neutrino-nucleus scattering (CE$\nu$NS) opens a new window to study the fundamental properties of neutrinos and allows to probe various scenarios of physics beyond the Standard Model (see \cite{barranco2005, lindner2017, billard2018} and references therein). \\
%
% Introduction to NUCLEUS with focus on Expt. Site & Background
The NUCLEUS experiment aims to employ gram-scale fiducial-volume cryogenic calorimeters with a demonstrated ultra-low threshold of E$_{th}$\,=\,19.7\,eV for nuclear recoils \cite{strauss2017}. The new experimental site designated for the installation of the NUCLEUS experiment is a 24\,m$^2$ basement room in an administrative building situated in between the two 4.25\,GW$_{th}$ reactor cores of the Chooz nuclear power plant (located in the Ardennes department in northeastern France). The site features an expected antineutrino flux as high as 2.1$\cdot$10$^{12}$\,$\bar{\nu}_e$/(s$\cdot$cm$^2$) with an average antineutrino energy as low as 1.5\,MeV \cite{angloher2019}. Besides the high rate and low energy that make nuclear reactors favorable neutrino sources for an accurate measurement of the CE$\nu$NS cross section, a low residual background rate in the target detectors is mandatory. Operating at an overburden of only 3\,m.w.e., muon-induced events are expected to be the dominant source of background \cite{heusser1995}. The muon-induced neutrons, which are particularly harmful due to the same experimental signature as the sought-for CE$\nu$NS (i.e. a nuclear recoil) can be reduced passively via moderation and absorption in polyethylene. Additionally, for an active background suppression of prompt muon-associated events and to reach the benchmark background index of 100\,counts\,/\,(keV kg day) in the sub-keV region \cite{angloher2019}, a muon veto with a detection efficiency of more than 99\,\% is required. \\
%
%NUCLEUS Muon Veto and necessity of Cryogenic Muon Veto
For the NUCLEUS experiment, the muon veto (shown schematically in Fig.\,1) will consist of a compact cube assembly of 28 single organic plastic scintillator panels equipped with wavelength shifting fibers and silicon photo multipliers (SiPM) to collect and detect the scintillation light. Each panel will be enclosed in a light-tight aluminium box and placed around the NUCLEUS target detector covering the largest possible part of its 4\,$\pi$ steradians of solid angle. The expected muon identification rate for such an arrangement of $\sim$\,460\,Hz implies an induced target detector dead time of a few percent. Inevitably, a large unshielded area remains in the upper side of the experiment, where the cryostat containing the NUCLEUS detector and its support structure will be inserted (see Fig.\,1). In order to close this gap without unnecessarily increasing the induced target detector dead time, a novel concept has been developed \cite{erhart2021}, consisting of a plastic scintillator based disc-shape active muon veto operating inside the NUCLEUS cryostat at sub-Kelvin temperatures. \\

\begin{figure}[htbp]
\begin{center}
\includegraphics[width=0.73\linewidth]{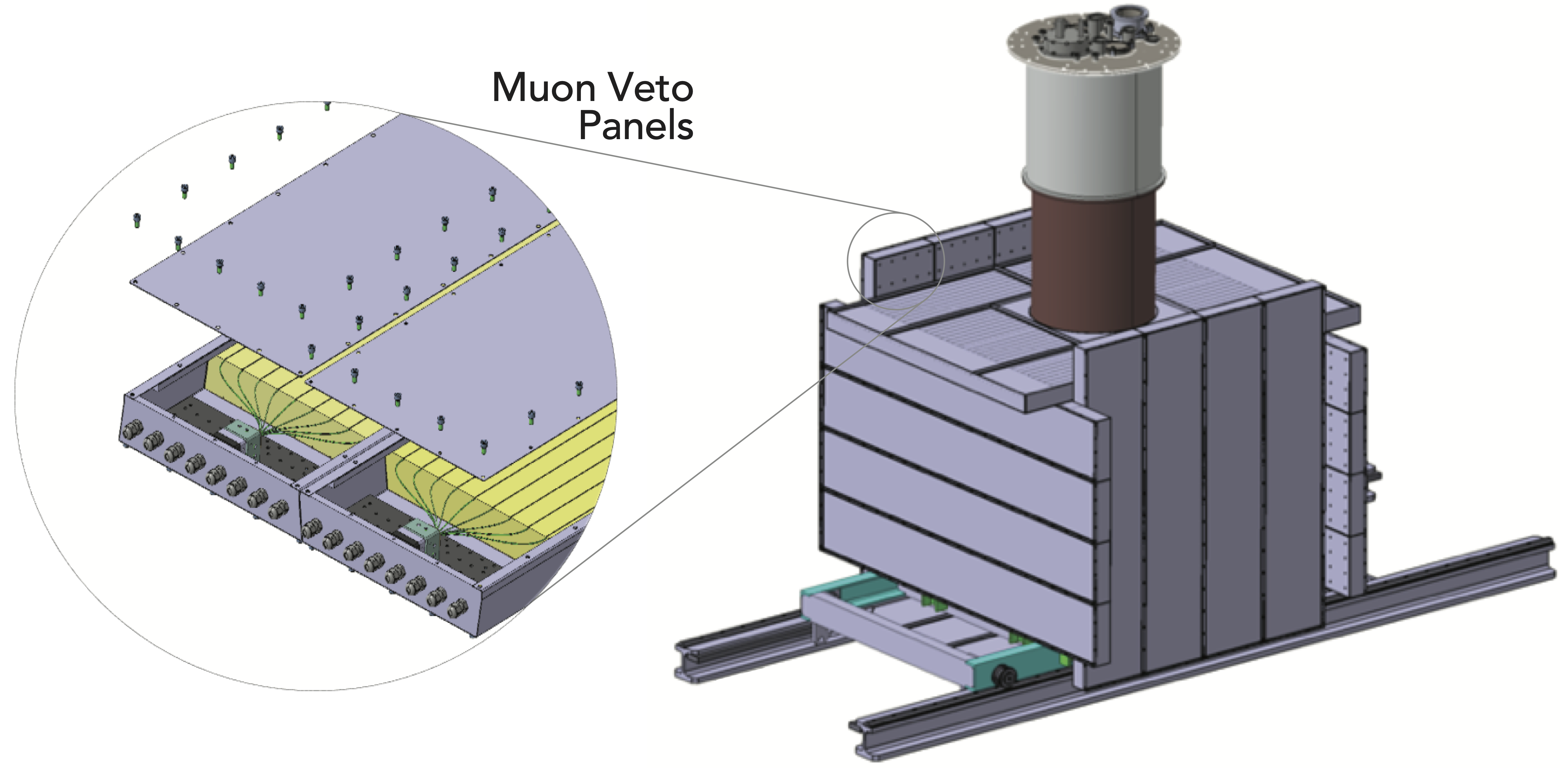}
\caption{Schematic drawing of the NUCLEUS experimental setup. The NUCLEUS cryostat containing the target detectors is shielded by several layers of active and passive materials. The outermost layer is an active muon veto, consisting of 28 single organic plastic scintillator panels each enclosed in a light-tight aluminium box and placed hermetically around the NUCLEUS cryostat. The single muon veto panels ({\it Left Zoom-In}) are equipped with wavelength shifting fibers and SiPMs to collect and detect the scintillation light. (Color figure online.)}
\end{center}
\label{fig1}
\end{figure}

\section{Detector Concept and Integration with the Cryogenic Infrastructure}
% Installation inside NUCLEUS cryostat -> shortly describe DDR
The NUCLEUS cryogenic muon veto will be installed inside a commercial LD 400 $^3$He\,/\,$^4$He dry dilution refrigerator provided by Bluefors \cite{LD400}. 
%
% Chosen detector concept, based on PS, describe design
The detector concept (shown schematically in Fig.\,2) consists - analogously to the NUCLEUS external muon veto concept - of plastic scintillator instrumented with a fiber based light-guide system together with a SiPM based read-out system. It is foreseen to install a disc of the polystyrene based plastic scintillator UPS 923-A \cite{UPS923-A} with an outer diameter of 297\,mm, a height of 50\,mm and a total mass of $\sim$\,3.6\,kg underneath the mixing chamber plate of the NUCLEUS cryostat. The disc features nine bended grooves, each of which guides a BCF-91A wavelength shifting fiber \cite{BCF-91A} through the interior of the plastic scintillator. The fibers are glued via two component clear epoxy resin into the grooves, assuring thorough fixation and maximization of the contact surface. An inner hole with a diameter of 45\,mm is required for the feed-through of the detector support structure.  
%
% Fiber guidance and SiPM attachment to 300K
The scintillation light is collected by the fibers and guided through the cryostat vessels towards a gain-stabilized and pre-amplified KETEK PE3325-WB TIA TP SiPM module \cite{PE3325-WB}, which is mounted on the bottom of the 300\,K plate inside the cryostat and thus operating at constant room temperature.
%
% Mechanical Support Structure as thermal coupling, Inner Shielding
The cryogenic muon veto constitutes together with a passive shielding (consisting of lead and polyethylene) the inner shielding of the NUCLEUS experiment. The mechanical holding structure of the inner shielding has been conceived such that a copper disc, connected via three Kevlar tie rods to the central tube of the cryostat, serves both as mechanical support structure and thermal bath for the plastic scintillator disc of the cryogenic muon veto. It is thermally coupled to the still stage of the cryostat, and thus thermalized at $\sim$\,850\,mK.

\begin{figure}[htbp]
\begin{center}
\includegraphics[width=0.93\linewidth]{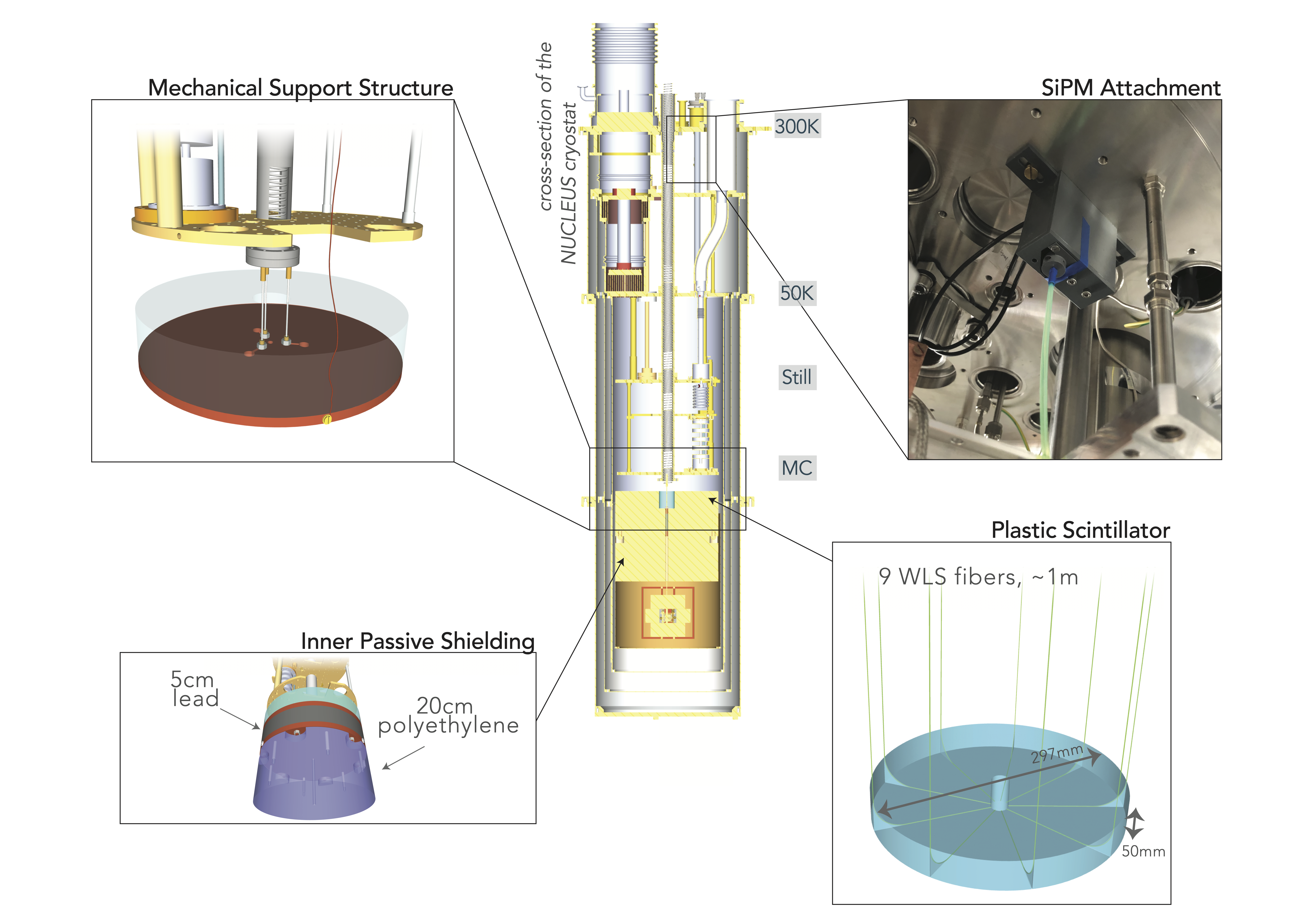}
\caption{Schematic drawing of the NUCLEUS cryogenic muon veto detector concept, with a cross-section of the NUCLEUS cryostat ({\it Center}) and zoom-in on the plastic scintillator disc ({\it Bottom Right}), the SiPM attachment to the 300\,K plate ({\it Top Right}), the inner passive shielding ({\it Bottom Left}) and the mechanical support structure ({\it Top Left}). (Color figure online.)}
\end{center}
\label{fig2}
\end{figure}

\section{Low Temperature Behavior of Organic Plastic Scintillators}

Given the scarcity of experimental studies on the thermalization behavior and the scintillation mechanism of organic plastic scintillators at sub-Kelvin temperatures, a proof-of-concept detector has been conceived in order to validate the feasibility of the intended NUCLEUS cryogenic muon veto detector concept. To this end, smaller-scale cylindrical pieces of plastic scintillators with a height and a diameter of 42\,mm each and four of the BCF-91A wavelength shifting fibers glued into small grooves on the side were assembled, mimicking the envisaged final design. These cryogenic muon veto prototypes are being operated within a customized copper holder on top of the still stage of the NUCLEUS cryostat and thermally coupled to a temperature of T$_{still}$\,$\sim$\,850\,mK. The temperature of the plastic scintillator can be monitored with a calibrated resistance thermometer on its top surface, and the light from the scintillator is guided through the fibers to a SiPM mounted at the bottom of the 300\,K plate of the cryostat.

\subsection{Thermalization Behavior of Organic Plastic Scintillators}
The thermalization behavior of organic plastic scintillators has been investigated for a sample of the polyvinyltoluene based plastic scintillator EJ-204 \cite{EJ-204}. The temperature curves of the still stage T$_{still}$ ({\it Blue}) and of the upper side of the cylindrical plastic scintillator T$_{PS}$ ({\it Red}) are shown in Fig.\,3. Starting from 100\,K, the still stage reaches its base temperature T$_{still}$\,$\approx$\,0.86\,K after $\sim$\,27\,h. The plastic scintillator sample fully thermalizes within $\sim$\,52\,h (with the time onset defined as the point in time in which T$_{still}$\,=\,100\,K), corresponding to a delay in thermalization time of the plastic scintillator w.r.t. the still stage of $\sim$\,25\,h. The cooling performance of the dry dilution refrigerator has not noticeably degraded due to the installation of the plastic scintillator piece.

\begin{figure}[htbp]
\begin{center}
\includegraphics[width=1\linewidth]{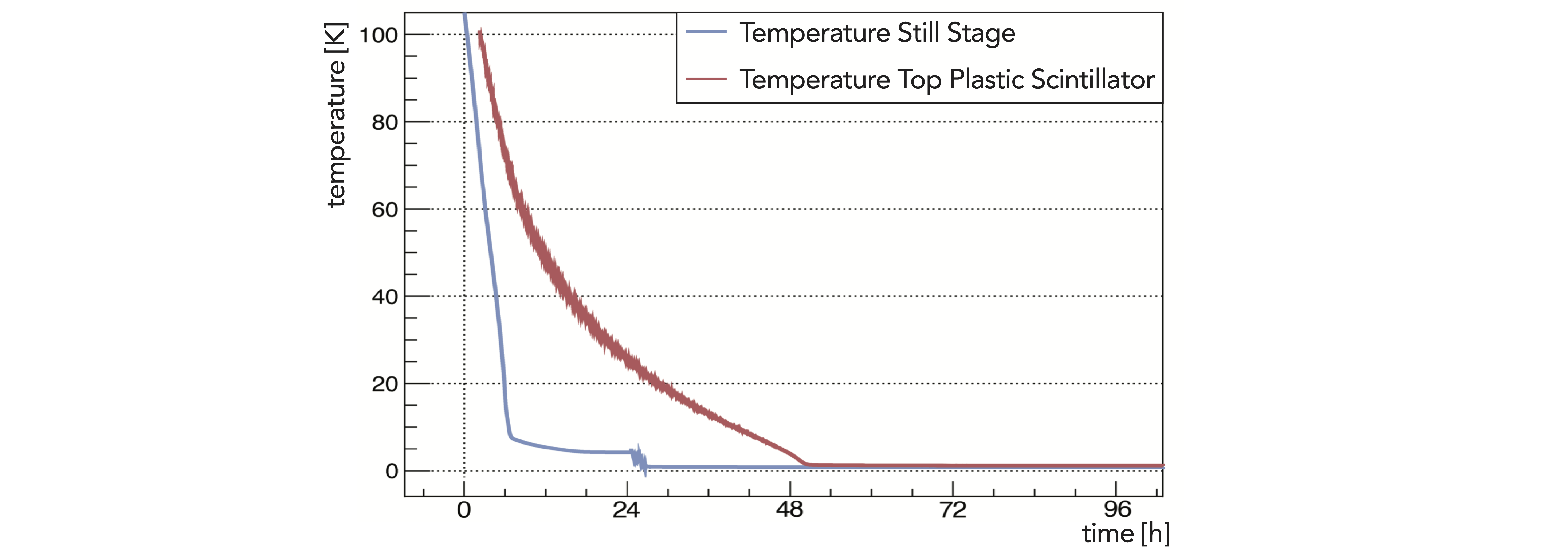}
\caption{Temperature curves of the upper side of the plastic scintillator prototype ({\it Red}) and of the still stage ({\it Blue}) during cool-down of the cryostat. The time onset is defined when the still stage reaches a temperature of T$_{still}$\,=\,100\,K. The still stage reaches its base temperature of T$_{still}$\,$\approx$\,0.86\,K within 27\,h. The thermal equilibrium temperature of the plastic scintillator T$_{PS}$\,$\approx$\,1.19\,K is reached within 52\,h. (Color figure online.)}
\end{center}
\label{fig3a}
\end{figure}

In summary, this observation states that the cooling of the NUCLEUS cryogenic muon veto to its operating temperature is achievable on a time scale completely reconcilable with the operational parameters envisaged for the NUCLEUS cryogenic infrastructure and is thus not expected to considerably affect them.
%When applying these observations to a prediction of the thermalization behavior of the larger-scale NUCLEUS cryogenic muon veto, an additional increase of thermalization time of $\sim$\,19\,\% is to be expected due to the difference in height.

\subsection{Temperature Dependence of the Muon Pulse Shape}
The temperature dependence of the pulse shape of a muon crossing through organic plastic scintillators has been studied for a sample of the polystyrene based plastic scintillator UPS 923-A. This resulted, to the best of the authors' knowledge, in the first reported measurements of organic plastic scintillators at sub-Kelvin temperatures, confirming the functionality of their principal scintillation process down to these temperatures. Two muon pulses with comparable amplitudes, captured at room temperature T$_{room}$\,$\approx$\,293\,K and at still temperature T$_{still}$\,$\approx$\,850\,mK, are exemplary shown in Fig.\,4. The pulse decay time of the scintillator, which is defined as the time after which the pulse has returned to 1/$e$ of its peak value, has been observed to increase from t$_{decay,\,room}$\,=\,47\,ns at room temperature to t$_{decay,\,still}$\,=\,64\,ns at still temperature. On average, the pulse decay time has been found to increase by $\sim$\,32\,\% towards lower temperatures.

\begin{figure}[htbp]
\begin{center}
\includegraphics[width=1\linewidth]{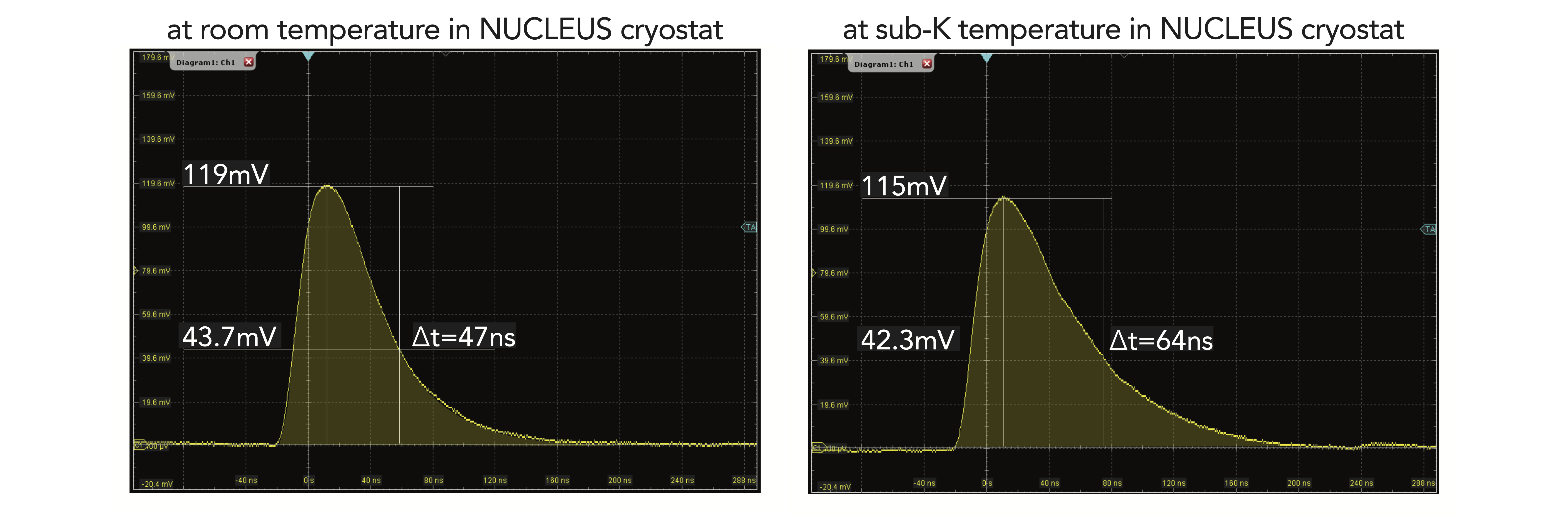}
\caption{Exemplary muon pulses recorded at room temperature T$_{room}$\,$\approx$\,293\,K ({\it Left}) and at still temperature T$_{still}$\,$\approx$\,850\,mK ({\it Right}). The pulses peak at values of 119\,mV and 115\,mV, respectively. The pulse decay time of the muon pulse increases from t$_{decay,\,room}$\,=\,47\,ns at room temperature to t$_{decay,\,still}$\,=\,64\,ns at still temperature. Averaged for a number of pulses, it can be observed that the pulse decay time is enhanced by 32\,\% at sub-Kelvin temperature. (Color figure online.)}
\end{center}
\label{fig3b}
\end{figure}

In the course of the featured measurements, the main components of the NUCLEUS cryogenic muon veto detector - namely the SiPM mounting in the inside of the cryostat, the associated fiber guidance and the plastic scintillator to be thermalized - have been installed and operated successfully. The practical operation of a muon veto based on an organic plastic scintillator inside a running dry dilution refrigerator has thus been demonstrated.

\section{Detector Commissioning at Room Temperature}
On the basis of these findings, the disc-shape plastic scintillator based muon veto has been assembled and commissioned at room temperature. Fig.\,5 shows the full charge spectrum recorded with the plastic scintillator disc ({\it Gray}), which features a pedestal, a region for low energetic background events (i.e. ambient gammas from natural radioactivity) and muon events following a Landau-like distribution. By operating the disc-shape scintillator in coincidence with an external scintillator panel beneath it, the charge spectrum can be restricted to exclusively muon events ({\it Red}). However, the coincidence criterion constrains the angular acceptance of impinging muons and hence their total rate. The conceived muon veto design allows unambiguous identification of the muon events and features sufficient discrimination capability between muons and ambient gammas.

\begin{figure}[htbp]
\begin{center}
\includegraphics[width=0.95\linewidth]{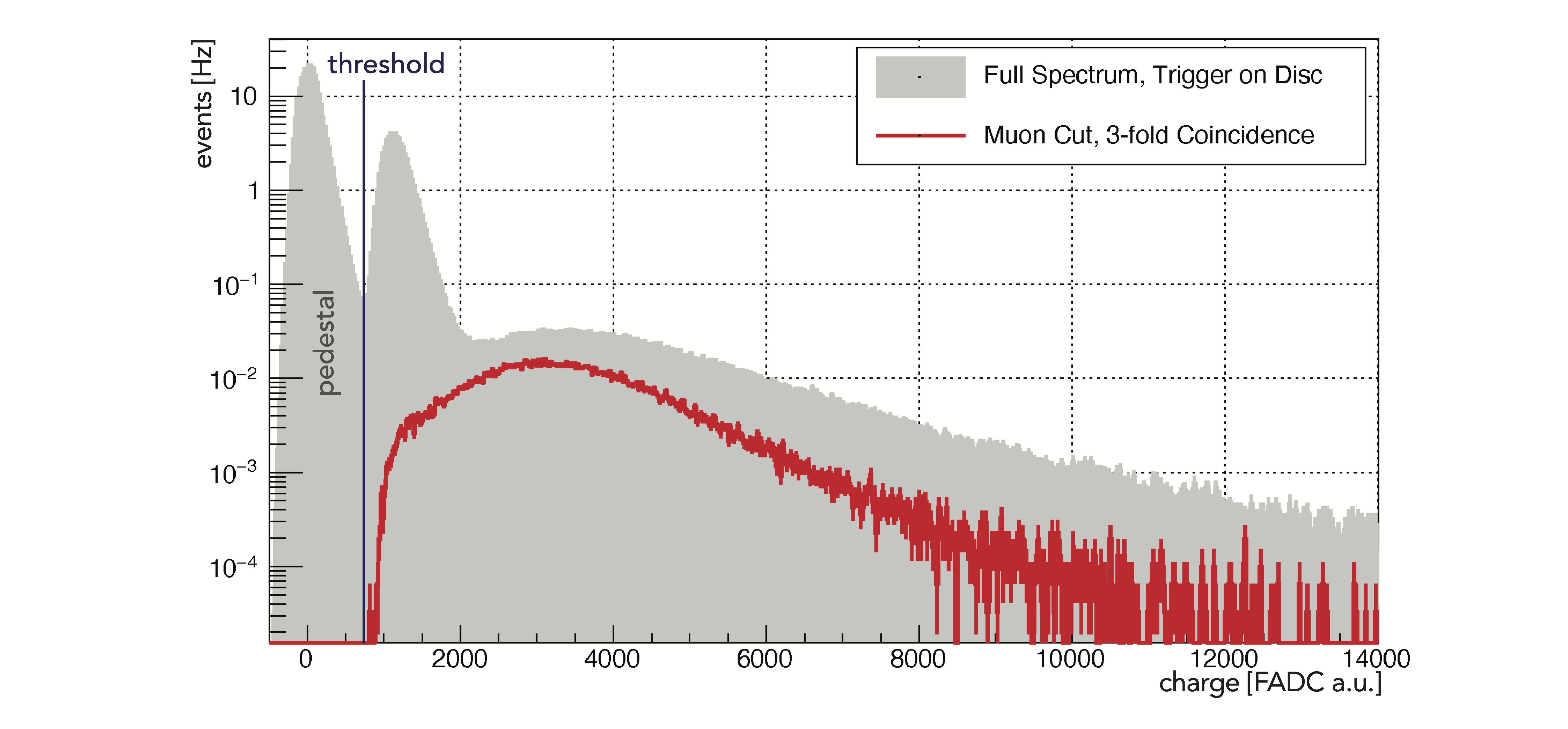}
\caption{Full charge spectrum recorded with the plastic scintillator disc ({\it Gray}) and restriction to exclusively muon events via coincidence criterion ({\it Red}). The full spectrum features a pedestal, a region for low energetic background events cut at a certain threshold and muon events following a Landau-like distribution. The mean value of the muon distribution can be determined to be $\sim$\,3063\,{\it FADC a.u.}, corresponding to muons impinging vertically with an average energy deposition in the plastic scintillator of around 10\,MeV. The restriction of the angular acceptance of muons leads to a decrease of their rate. (Color figure online.)}
\end{center}
\label{fig4}
\end{figure}

\section{Conclusion and Outlook}
By investigating the thermalization behavior of organic plastic scintillators and the temperature dependence of the muon pulse shape, the foundations for the development of an organic plastic scintillator based muon veto operating at sub-Kelvin temperatures inside the NUCLEUS cryostat have been laid. Within the framework of the work presented, the intended detector concept - i.e. a plastic scintillator based detector equipped with a fiber-based light-guide system together with a SiPM-based read-out system - has been proven for its applicability inside a running dry dilution refrigerator and commissioned at room temperature. 
A detailed characterization of the disc on its efficiency, its capability to separate muons and gammas, and its light yield homogeneity is ongoing. Furthermore, the phenomenology of organic plastic scintillators at different temperatures, in particular possible temperature dependencies of the detectors' light output, is currently being investigated within the established experimental setup. The first deployment of the cryogenic muon veto in coincidence with the external muon veto is planned for the commissioning of the NUCLEUS experiment, which will be carried out at Technical University of Munich in 2022-23.

\begin{acknowledgements}
We thank Maurice Chapellier and Patrick Champion for their support and advice during initial prototype tests at DPhN ({\it IRFU, CEA, Universit\'{e} Paris Saclay}). This research has been supported by the ERC-StG2018-804228 ”NUCLEUS”, by which the NUCLEUS experiment is funded and the SFB1258 "Neutrinos and Dark Matter in Astro- and Particle Physics".
\end{acknowledgements}

\pagebreak

\end{document}